\documentclass[amsmath,amssymb,aps,prl,showpacs,twocolumn]{revtex4}

\begin{document}
\title{ Cosmological birefringence constraints on light}

\author{Qasem Exirifard}
\affiliation{Physics Department, Institute for Advanced Studies in Basic Sciences, P.O.Box 45195-1159, Zanjan, Iran}
\affiliation{School of Physics, Institute for Research in Fundamental Sciences, P.O.Box 19395-5531, Tehran, Iran}
\email{exir@theory.ipm.ac.ir}

\begin{abstract} 
 We calculate the birefringence in the vacuum for light at the leading and sub-leading orders for the CPT-even part of the SME. We report that all the LIV coefficients absent  in the leading order, but the isotropic one, contributes to the sub-leading order birefringence. We consider models free of the first order birefringence. We then show that infrared, optical, and ultraviolet spectropolarimetry of cosmological sources bound the LIV coefficients  to less than $ 10^{-16}$. This improves  the best current bound on the parity-odd coefficients by two orders  of magnitude and establishes  the isotropy of the one-way light speed with the  precision of $41 \frac{nm}{s}$. 
\end{abstract}
\preprint{IPM/P-2010/034}

\maketitle

SME \cite{SME0}, the most general Extension of the Standard Model of elementary particles that includes all the parameters  breaking the Lorentz invariance, provides a framework to search for Lorentz Invariance Violating  (LIV) terms in all the sectors of the standard model. So LIV terms in the SME model, in contrary to the Robertson-Mansouri-Sexl Model \cite{Robertson,Mansouri}, are bounded by various phenomena not only the velocity of the light and clock synchronization  measurements \cite{SME}.  This has stemmed  recent research aimed to detect or study the LIV terms in various fields, including the  classical solutions of SME electrodynamics \cite{Bailey:2004na,Kobakhidze:2007iz, Casana:2008sd} , radiation spectrum of the electromagnetic waves and CMB data \cite{Kahniashvili:2008va,Lue:1998mq,Feng:2006dp,Wu:2008qb,Kostelecky:2007zz,Kamionkowski:2008fp,Balaji:2003sw},  black body radiation in finite temperature in SME electrodynamics \cite{Casana:2009xf,Casana:2008ry,Fonseca:2008xa,Casana:2009dq},  LIV terms in higher dimensional scenarios \cite{Gomes:2009ch,Baukh:2007xy},  synchrotron radiation \cite{Bocquet:2010ke, Gurzadyan:2010bt}, Cherenkov radiation \cite{Altschul:2006zz,Kaufhold:2007qd,Klinkhamer:2007ak} and modern cavity resonators or interferometry experiments \cite{Eisele:2009zz,Muller:2007zz,Tobar:2009gw,NaturePhysics,Saathoff:2003zz,Carone:2006tx, Hohensee:2010an}.  

Among various constraints imposed on the LIV terms of SME model, however, the absence of cosmological birefringence sets the most stringent constraint on the 10 out of 19 dimensionless parameters of the CPT-even part of the pure mSME electromagnetic sector \cite{Kostelecky:2001mb,Kostelecky:2002hh}: these ten parameters should be smaller than $2\times 10^{-32}$.  Five of the nine remaining parameters ($\widetilde{\kappa}_{e-}$) are parity-even, 3 ($\widetilde{\kappa}_{o+}$) are parity-odd and one ($\kappa_{tr}$) is an isotropic LIV term. The parity-even terms are constraint to be less than $10^{-17}$ by the most recent  Michelson-Morley-type experiment \cite{Eisele:2009zz}. Considering the motion of earth around the Sun and the fact that boost mixes various LIV parameters,  \cite{Eisele:2009zz} also requires  the parity-odd parameters to be less than $10^{-13}$. Ref. \cite{Bocquet:2010ke, Gurzadyan:2010bt} uses the absence of sidereal variations in the energy of Compton-edge photons at the ESRF's GRAAL facility,  to set the limit of  $10^{-14}$ on the parity odd parameters.  So far, no experiment or observation (in low energy physics) has bounded the parity odd parameters beyond $10^{-14}$.  So perhaps it is interesting to refine/translate some current data into stronger bounds on the parity odd coefficients, the parameters that encode the anisotropy in the one-way light speed.

We consider CPT-even part of the pure SME electromagnetic sector.  We obtain the birefringence in the vacuum at the leading and the sub-leading orders. We show that all the  coefficients absent  in the leading order but the isotropic one contribute to the birefringence  at the sub-leading order. This means that $\tilde{\kappa}_{o+}$ and $\tilde{\kappa}_{e-}$  must not be called non-birefringent terms, they do contribute to the birefringence, a fact that has been noted also in \cite{Casana:2009xf,Casana:2009dq} . So in  models which are free of the first order birefringence, absence of the cosmological birefringent indeed bounds $\tilde{\kappa}_{o+}$ and $\tilde{\kappa}_{e-}$. These models include SME camouflage models \cite{Kostelecky:2009zp}.  We consider these models and we show that  the absence of the cosmological birefringence  bounds each element of the $\tilde{\kappa}_{o+}$ to less than $8 \times 10^{-17}$ at $90\%$ confidence level, or equivalently the squared-sum of its three elements to $1.4 \times 10^{-16}$ at $90\%$ confidence level.  This establishes  the isotropy of the one-way light speed with the  precision of $41 \frac{nm}{s}$.  This precision is two order of magnitudes better than the best limit, as we shall show by reviewing the literature.

\section{Sub-leading order birefringence}  
A Lorentz violating extension of the pure massless $U(1)$ gauge sector of the standard model reads
\begin{eqnarray}
{\cal L} &=& -\frac{1}{4} F_{\mu\nu} F^{\mu\nu}\,-\,\frac{1}{4} (k_{F})_{\mu\nu\lambda\eta} F^{k\lambda} F^{\mu\nu}\nonumber\\ 
&&+ \frac{1}{2} (k_{AF})^k \epsilon_{k\lambda\mu\nu} A^\lambda F^{\mu\nu} ,\label{MGE}
\end{eqnarray}
Eq. \eqref{MGE} is the most general extension of QED which is quadratic in the gauge field, and contains no more than two derivatives\footnote{The behavior of photons in the presence of Lorentz and CPT violation operator with arbitrary (or other) mass dimension is studied in \cite{Kostelecky:2009zp, 
Mewes:2010ig, Gubitosi:2010dj}. } and does not respect the Lorentz symmetry. $(k_{AF})^k$ represents a CPT-odd Lorentz violating term. This term is vanishing for theoretical reasons  \cite{Colladay:1996iz, Kostelecky:2000mm} and cosmological birefringence requires it to be smaller than $10^{-42} GeV$ \cite{Carroll:1989vb, Jackiw:1999yp, Wardle:1997gu}. Here we would like to address the constraints on $k_{F}$. The theoretical consistency of the parity-even terms are studied in \cite{Casana:2009xs,Casana:2010nd}. We, thus, set $k_{AF}\,=\, 0$, and consider\footnote{ Here $k_F$ is independent of the space-time. In more general contexts  \cite{Kostelecky:2003fs, Bailey:2006fd,Kostelecky:2005ic}, it may not.} 
\begin{equation}\label{MGEh}
{\cal L} \,=\, -\frac{1}{4} F_{\mu\nu} F^{\mu\nu}\,-\frac{1}{4} (k_{F})_{\mu\nu\lambda\eta} F^{k\lambda} F^{\mu\nu}.
\end{equation}
$(k_{F})_{\mu\nu\lambda\eta}$ has the symmetries of the Riemann tensor, so only $20$ out of its $256$ components are algebraically independent. Its double trace should be zero.   So only $19$ algebraically independent components of the $(k_{F})_{\mu\nu\lambda\eta}$ contribute to the equations of motion of the gauge field. Ref. \cite{Kostelecky:2002hh} introduces an interesting re-parametrization of these components by enclosing them in a parity-even and parity-odd subsectors, respectively  $\tilde{k}_e$ and $\tilde{k}_o$:
\begin{align}
\left(  \widetilde{\kappa}_{e+}\right)  ^{jk} &  =\frac{1}{2}(\kappa_{DE}+\kappa_{HB})^{jk}, ~~\kappa_{\text{tr}}=\frac{1}{3}\text{tr}(\kappa_{DE}),\\
\left(  \widetilde{\kappa}_{e-}\right)  ^{jk} &=\frac{1}{2}(\kappa_{DE}-\kappa_{HB})^{jk}-\frac{1}{3}\delta^{jk}(\kappa_{DE})^{ii},\\
\left(  \widetilde{\kappa}_{o+}\right)  ^{jk} &  =\frac{1}{2}(\kappa_{DB}+\kappa_{HE})^{jk},\\
\left(  \widetilde{\kappa}_{o-}\right)^{jk} &=\frac{1}{2}(\kappa_{DB}-\kappa_{HE})^{jk}~~.
\end{align}
The $3\times3$ matrices $\kappa_{DE},\kappa_{HB},\kappa_{DB},\kappa_{HE} $ are
given as:
\begin{align}
\left(  \kappa_{DE}\right)  ^{jk} &  =-2(k_F)^{0j0k},\text{ }\left(  \kappa_{HB}\right)  ^{jk}=\frac{1}{2}\epsilon^{jpq}\epsilon^{klm}(k_F)^{pqlm}%
\nonumber\\
\left(  \kappa_{DB}\right)  ^{jk} &  =-\left(  \kappa_{HE}\right)^{kj}=\epsilon^{kpq}(k_F)^{0jpq}.\label{P2}%
\end{align}
Note that $\widetilde{\kappa}_{e+},\widetilde{\kappa}_{e-},\widetilde{\kappa}_{o-}$ are  traceless and symmetric while $\widetilde{\kappa}_{o+}$ is  anti-symetric. $\kappa_{\text{tr}}$ is a number, it represents the isometric LIV term. In term of this parametrization, the Lagrangian density reads
\begin{eqnarray}
\mathcal{L}&=&
\frac{1}{2}\left[  \left(  1+\kappa_{\text{tr}}\right)
\mathbf{E}^{2}-\left(  1-\kappa_{\text{tr}}\right)  \mathbf{B}^{2}\right] \nonumber\\
&&+\frac{1}{2}\mathbf{E}\cdot\left(  \widetilde{\kappa}_{e+}+\widetilde{\kappa
}_{e-}\right)  \cdot\mathbf{E}
  -\frac{1}{2}\mathbf{B}\cdot\left(  \widetilde{\kappa}_{e+}-\widetilde
{\kappa}_{e-}\right)  \cdot\mathbf{B}\nonumber 
\\&&+\mathbf{E}\cdot\left(  \widetilde
{\kappa}_{o+}+\widetilde{\kappa}_{o-}\right)  \cdot\mathbf{B}~~\,,
\end{eqnarray}
where $\mathbf{E}$ and $\mathbf{B}$ respectively are the electric and magnetic field. This new parametrization  clearly illustrates the analogy between the propagation of light in the vacuum of the theory with the propagation of light in a general anisotropic media,  a field intensely explored in optics.  For the moment we consider \eqref{MGEh}. The first variation of \eqref{MGEh} with respect to $A_{\mu}$ gives its equation of motion:
\begin{equation}
\partial_{\alpha} F_{\mu}^{~\alpha} + (k_F)_{\mu\alpha\beta\gamma} \partial^{\alpha} F^{\beta\gamma}\,=\,0\,,
\end{equation}
which is supplemented with the usual homogeneous Maxwell equation:
\begin{equation}
\partial_\mu \widetilde{F}^{\mu\nu}\,=\,0.
\end{equation}
These equations for a plane electromagnetic wave with wave 4-vector $p^{\alpha}= (p^0, \vec{p})$, $F_{\mu\nu}= F_{\mu\nu}(p) e^{-i p_{\alpha} x^{\alpha}} $,   lead to the modified Ampere law \cite{Kostelecky:2001mb}:
\begin{equation}
M^{jk} E^{k} \equiv (-\delta^{jk} p^2 -p^{j}p^{k}- 2 (k_F)^{j\beta\gamma k} p_{\beta} p_{\gamma}) E^{k} \,=\, 0
\end{equation}
which has non-trivial solution for the electric field provided that the $M$ matrix has zero eigenvalues. The zero eigenvalues of  $M^{jk}$ give the dispersion relation in the vacuum. In order to obtain these zero eigenvalues let a prime coordinate be considered where in $\tilde{p}^{\alpha}= (p^0, 0,0, p^3)$. In the prime coordinate, the zero eigenvalues at the leading order reads
\begin{eqnarray}\label{BiF}
p_\pm^{0} &=& p^3 (1 + \frac{k_{11}+ k_{22}}{2}\pm \sqrt{k_{12}^2 +\frac{(k_{11}-k_{22})^2}{4}}),\\
p_+^{0}& -& p_{-}^{0} \, = \,2 p^3 \sqrt{k_{12}^2 +\frac{(k_{11}-k_{22})^2}{4}}\,, \label{Bi1}
\end{eqnarray}
where 
\begin{equation}
k_{ij}= (\tilde{k}_F)_{i \alpha j\beta}\, \frac{\tilde{p}^{\mu} \tilde{p}^{\nu}}{|p_3|^2}\,,
\end{equation}
wherein $(\tilde{k}_F)_{i \alpha j\beta}$ represents the component of the $k_F$ tensor in the prime coordinate.  
Eq. \eqref{BiF} shows that at the leading order only two combinations of the components of the k-matrix contribute to the birefringence.  From $19$ algebraically independent components of the $(k_F)_{\mu\nu\lambda\eta}$ only $10$ of them contributes to the $k_{11}-k_{22}$ and $k_{12}$ for an arbitrary four-wave vector: $p^{\alpha}= (p^0, \vec{p})$. An acceptable choice of these ten combinations is: 
\begin{eqnarray}
k^a_1 &=& \bigl( 
(k_F)^{0213},~
(k_F)^{0123},~
\\
&&
(k_F)^{0202}-(k_F)^{1313},~
(k_F)^{0303}-(k_F)^{1212},~
\nonumber\\
&&
(k_F)^{0102}+(k_F)^{1323},~
(k_F)^{0103}-(k_F)^{1223},~
\nonumber\\
&&
(k_F)^{0203}+(k_F)^{1213},~
(k_F)^{0112}+(k_F)^{0323},~
\nonumber\\
&&
(k_F)^{0113}-(k_F)^{0223},~
(k_F)^{0212}-(k_F)^{0313}
\bigr).
\label{ka}\nonumber
\end{eqnarray}
Note that elements in $k^a_1$ are contained in the matrices $\tilde{\kappa}_{e+}$ and $\tilde{\kappa}_{o-}$ \cite{Kostelecky:2001mb,Kostelecky:2002hh}.
Ref. \cite{Kostelecky:2001mb,Kostelecky:2002hh} use infrared, optical, and ultraviolet spectropolarimetry of various cosmological sources at distances  $0.04-2.08 Gpc$ \cite{data1,data2,data3,data4,data5,data6,data9,data10} and bound the components of $\tilde{\kappa}_{e+}$ and $\tilde{\kappa}_{o-}$ to less than $2 \times 10^{-32}$ at $90\%$ confidence level. Some combinations of $k_a$  are further restricted to less than $10^{-37}$ using linear polarization  data of gamma rays of cosmological sources \cite{Kostelecky:2006ta}.  Optical and microwave cavities  can measure the  components of $k_F$ that does not contribute to the linear birefringence with  the precision of  $10^{-9}-10^{-16}$ \cite{Kostelecky:2002hh}.

We note that the best precision achieved in the laboratories is about or less than the square root of the precision  of the cosmological constraints on the leading contribution of the LIV terms to the birefringence in the vacuum.  So  the sub-leading contribution of the  coefficients  having no contribution at the leading order, can not be neglected. Ref. \cite{Kostelecky:2001mb,Kostelecky:2002hh}  tacitly presumes  that all the coefficients are at the same order of magnitude and provides the limit on \eqref{ka}. When the coefficients are not at the same order, the absence of birefringence leads to ten non-linear inequalities among the nineteen parameters \footnote{An elegant approach to derive the exact birefringence relation is given by \cite{Kostelecky:2009zp}, however, sub-leading  order contributions have not been calculated.}. In this note, we assume that ``the leading order birefringence is zero for some theoretical reasons'' and we calculate the second order birefringence. The models which have zero order birefringence includes camouflage model given in Table XVIII of  ref. \cite{Kostelecky:2009zp}. 

In the models we consider, we have $k^a_1=\tilde{\kappa}_{e+}=\tilde{\kappa}_{o-}=0$. This means that in  the prime coordinate $k_{11}-k_{22}=k_{12}=0$. In the prime coordinate the zero eigenvalues of the M-Matrix at the sub-leading order then yields:
\begin{equation}\label{Bi2}
p_+^{0}-p_{-}^{0} \,=\, 2 p^3 (k_{13}^2+ k_{23}^2)
\end{equation}
where
\begin{eqnarray}
k_{13} \,p_3^2 & =& (\tilde{k}_F)^{1030} p_0^2 + (\tilde{k}_F)^{1330} p_0 p_3\,,\\
k_{23}\, p_3^2& =& (\tilde{k}_F)^{2030} p_0^2 + (\tilde{k}_F)^{2330} p_0 p_3\,.
\end{eqnarray}
Only eight combinations of the nine remaining coefficients  contributes to the $k_{13}$ and $k_{23}$  for an arbitrary four-wave vector. An acceptable choice for these is
\begin{eqnarray}
k^a_2 &=& \bigl( 
(k_F)^{0102},(k_F)^{0103},(k_F)^{0203},(k_F)^{0202}-(k_F)^{0101},
\nonumber\\
&&\qquad
(k_F)^{0303}- (k_F)^{0202},~
(k_F)^{0112}-(k_F)^{0323},
\nonumber\\
&&
(k_F)^{0113}+(k_F)^{0223},~
(k_F)^{0212}+(k_F)^{0313}
\bigr).
\label{ka2}
\end{eqnarray}
Let it be outlined that only the isotropic LIV term, $\kappa_{tr}$,  is missing in  \eqref{ka2}. At the sub-sub-leading order, in the prime coordinate also $k_{33}$ contributes to the birefringence. So the isotropic   LIV term contributes to the birefringence at the sub-sub-leading order.  Also note that no  element in \eqref{ka2} can be written as a combination of the elements in \eqref{ka} and the vanishing of  double trace of $k_F$.

Comparing \eqref{Bi1} with \eqref{Bi2} leads to the conclusion that $ (k_{13})^2$ is experimentally constraint as so much as is $k_{12}$. This implies that  the squared of the elements in \eqref{ka2} are constraint as so much as the elements in \eqref{ka}. Ref. \cite{Kostelecky:2002hh,Kostelecky:2001mb}  bound  the square-averaged of all the terms in \eqref{ka}, $\sqrt{(\sum_{i=1}^{10}k_1^i)^2}$, to less than $2 \times 10^{-32}$ at 90\% confidence level. So each element in \eqref{ka} is bound to less than $\frac{2}{\sqrt{10}}\times 10^{-32}$. Subsequently each element in \eqref{ka2} is bound less than $\sqrt{\frac{2}{\sqrt{10}}}\times 10^{-16}=8 \times 10^{-17}$.  The parity even terms in \eqref{ka2} are constraint to  less than $10^{-17}$ by \cite{Eisele:2009zz}. This let us set the parity even coefficients in \eqref{ka2} to zero at the precision we are working in. This yields that the square sum of the parity-odd coefficients in \eqref{ka2},  $\sqrt{(\sum_{i=1}^{3}k_2^i)^2}$, is less than  $1.4 \times 10^{-16}$ at the conservative 90\% confidence level. This proves that the one way light speed is isotropic with the precision of $41 \frac{nm}{s}$.

Two combinations of terms in \eqref{ka} are further bounded less than $10^{-37}$ by the analyze of \cite{Kostelecky:2006ta,Kahniashvili:2006dt} on the gamma rays from GRB 930131 and GRB 960924 \cite{data11}.  Repeating the analyze of \cite{Kostelecky:2006ta} for  other gamma ray sources can measure all the components of \eqref{ka} with the precision of  $10^{-37}$, and subsequently in theoretical models wherein $k_1^a=0$ all the components of \eqref{ka2} with the precision of about $ 10^{-19}$.

The most recent  Michelson-Morley-type experiment \cite{Eisele:2009zz}, improving previous bounds \cite{Muller:2007zz},  reports the bound of $10^{-13}$  on the parity odd coefficients in \eqref{ka2}.   Searching for compton-edge photons at the ESRF's GRAAL facility \cite{Bocquet:2010ke, Gurzadyan:2010bt} improves this precision by one order.  So the bound we have provided improves the best low energy precision on the one way light speed isotropy by two orders of magnitude.

Ref. \cite{Altschul:2009xh} considers the synchrotron emission rate of fast moving electrons and positrons  in LEP and the measurements performed at the Z pole energy of 91 GeV,  and concludes the limit of  $|\kappa_{tr}| < 5 \times 10^{-15}$. This is orders of magnitude improvement on the previous bounds on $\kappa_{tr}$ \cite{Eisele:2009zz,Muller:2007zz,Tobar:2009gw,NaturePhysics,Saathoff:2003zz,Carone:2006tx, Hohensee:2009zk,Hohensee:2010an}.  Considering the motion of earth around the Sun, our limit of $8\times 10^{-17}$ on each elements of $\kappa_{o+}$ implies a double sided bound of $|\kappa_{tr}| < 8 \times 10^{-13}$.  But this limit is apparently weaker than that  of ref. \cite{Altschul:2009xh}. So our improvement on the limit of $\kappa_{o+}$ does not lead to the improvement of the best current limit on $\kappa_{tr}$. 
 
Ref. \cite{Altschul:2006zz,Kaufhold:2007qd} propose that ultrahigh-energy cosmic rays (UHECRs) have the potential to place further limits on all the non-birefringent parameters by the inferred absence of vacuum Cherenkov radiation.  Ref. \cite{Klinkhamer:2007ak}, having inferred   the absence of Cherenkov radiation for 29 UHECRs at energy scale of $10^{10}GeV$ \cite{Abraham:2004dt},  states bound of $10^{-18}$ on the nine non-birefringent terms. The contribution of the massive  LIV terms \cite{Kostelecky:2009zp, Mewes:2010ig, Gubitosi:2010dj}, however,  necessarily can not be ignored for the energy scale of the ref. \cite{Klinkhamer:2007ak}. Here, we are providing  bounds on the parameters at low energies,  energy scales that the contributions of the massive LIV terms can be neglected. Our results combined with \cite{Altschul:2006zz,Kaufhold:2007qd} prove that up to the scale of $10^{10} GeV$, no new LIV terms -in addition to the low energy ones- is dynamically generated. So the fundamental scale of quantum gravity likely should be higher than $10^{10} GeV$ should quantum gravity  be resolved through break of the Lorentz invariance \footnote{Also note that the form of the modifications to the dispersion relations in Lorentz Voilated quantum gravity are restricted \cite{Schuller}.}.

There exist proposals to measure parity-odd LIV terms by electrostatics or magnetostatics systems \cite{Kobakhidze:2007iz,Casana:2008sd}.   Though these experiments are remained to be implemented, the bound  we provide achieves their precision.  Ref. \cite{Exirifard:2010xp} proposes a triangular Fabry-Perot resonator based on the Trimmer experiment \cite{Trimmer:1973nn}, using which in the current resonator experiments \cite{Eisele:2009zz,Muller:2007zz} would improve the bound we report on the parity-odd parameters. This experiment is the only proposed experiment that would improve our bound, however, it is remained to be implemented. 

\textit{\small{I thank V. A.  Kostelecky for very helpful  comments.}} 

\providecommand{\href}[2]{#2}\begingroup\raggedright

\end{document}